\documentclass[journal=jpclcd,manuscript=letter]{achemso}
\setkeys{acs}{articletitle = true}
\usepackage[version=3]{mhchem} 
\usepackage{natbib}
\usepackage{times}
\usepackage{achemso}
\usepackage{amssymb,amsfonts,amsmath}

\usepackage{graphicx}
\newcommand{\edit}[1]{{{#1}}}

\usepackage{color}

\author {Yan Yan}
\affiliation{Department of Chemistry, State University of New York at Buffalo, Buffalo, NY 14260-3000, USA}
\altaffiliation{School of Sciences, Changchun University, Changchun, 130022, China}
\author{Tiange Bi}
\affiliation{Department of Chemistry, State University of New York at Buffalo, Buffalo, NY 14260-3000, USA}
\author{Nisha Geng}
\affiliation{Department of Chemistry, State University of New York at Buffalo, Buffalo, NY 14260-3000, USA}
\author{Xiaoyu Wang}
\affiliation{Department of Chemistry, State University of New York at Buffalo, Buffalo, NY 14260-3000, USA}
\author{Eva Zurek}
\email{ezurek@buffalo.edu}
\affiliation{Department of Chemistry, State University of New York at Buffalo, Buffalo, NY 14260-3000, USA}

\title{A Metastable CaSH$_3$ Phase Composed of HS Honeycomb Sheets that is Superconducting Under Pressure}

\begin{document}

\begin{center}
\textbf{Abstract}  
\end{center}

Evolutionary searches predicted a number of ternary phases that could be synthesized at pressures of 100-300~GPa. $P6_3/mmc$ CaSH$_2$, $Pnma$ CaSH$_2$, $Cmc2_1$ CaSH$_6$, and $I\bar{4}$ CaSH$_{20}$ were composed of a Ca-S lattice along with H$_2$ molecules coordinated in a ``side-on'' fashion to Ca. The H-H bond lengths in these semiconducting phases were elongated because of H$_2$ $\sigma\rightarrow$ Ca d donation, and Ca d$\rightarrow$ H$_2$ $\sigma^*$ back-donation, via a Kubas-like mechanism. $P\bar{6}m2$ CaSH$_3$, consisting of two-dimensional HS and CaH$_2$ sheets, was metastable and metallic above \edit{128~GPa}. The presence of van Hove singularities increased its density of states at the Fermi level, and concomitantly the superconducting critical temperature, which was estimated to be as high as \edit{$\sim$100~K at 128~GPa}. This work will inspire the search for superconductivity in materials based upon \edit{honeycomb HX (X=S, Se, Te), and MH$_2$ (M=Mg, Ca, Sr, Ba) layers} under pressure. \\[2ex]

\noindent \textbf{TOC Graphic}

\begin{figure*}
\begin{center}
\includegraphics[width=5.0cm]{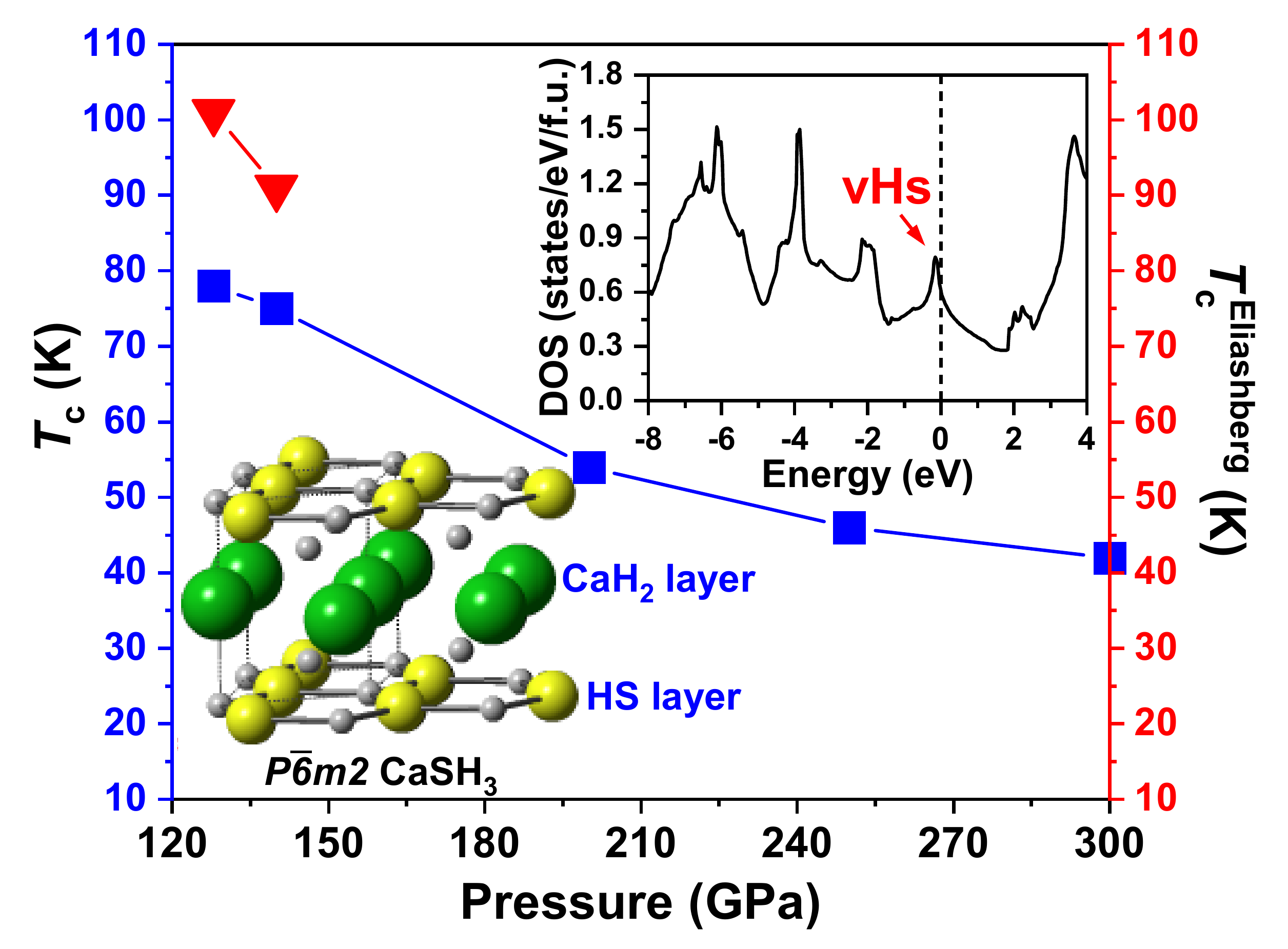}
\end{center}
\end{figure*}

\newpage

The last years have witnessed tremendous advances in the search for conventional superconductivity in high pressure hydrides \cite{Struzhkin:2015a,Zurek:2016d,Zurek:2016j,Duan:2017a,Wang:2017a,Pickard:2019-Review,Livas:2020-Review,Zurek:2018d,Semenok:2020a}. First-principles based crystal structure prediction (CSP) techniques have pinpointed two classes of binary hydrides as being the most promising for high temperature superconductivity \cite{Zurek:2018m}. The first are composed of $p$-block elements in groups 13-16, such as the hydrides of phosphorus \cite{Zurek:2015j,Zurek:2017c,Liu:2016-P,Flores:2016-P}, and sulfur \cite{Cui:2014a,Yao-S-review:2018}. The second are alkaline or rare earth metal hydrides such as CaH$_6$ \cite{wang2012superconductive}, which was predicted to have a  superconducting critical temperature,  $T_c$, of 220-235~K at 150~GPa. At the same time experiments have measured record breaking $T_c$ values, first in $Im\bar{3}m$ H$_3$S ($T_c$ of 203~K near 150~GPa)  \cite{drozdov2015conventional} , and later in an $Fm\bar{3}m$  symmetry LaH$_{10}$ phase ($T_c$ of 250-260~K near 200~GPa \cite{somayazulu2019evidence, drozdov2019superconductivity}) that was predicted prior to its synthesis \cite{liu2017potential,peng2017hydrogen}.

Ternary hydrides are the next frontier. Experimentally, only a few systems with low $T_c$s have been studied \cite{Muramatsu:2015,Meng:2019}. The theoretical prediction of ternaries poses considerable challenges because of the combinatorial complexity, wide stoichiometry range, and potentially large unit cell sizes. At the same time, the vast parameter space tantalizes at the prospect of discovering a stable or metastable species that has a high $T_c$ at mild pressures. To date computations have predicted only a handful of ternaries with $T_c$s larger than 100~K, including LiPH$_6$ (167~K at 200~GPa) \cite{Shao:2019}, LiP$_2$H$_{14}$ (169~K at 230~GPa) \cite{Li:2020}, Li$_2$MgH$_{16}$ (473~K at 250~GPa) \cite{Sun:2019}, CH$_4$-SH$_3$ (194~K at 150~GPa \cite{Zurek:2020b}, or 181~K at 100~GPa \cite{Sun:2020a}), SH$_3$-SeH$_3$ (196~K at 200~GPa) \cite{Liu:2018}, MgCH$_4$ (121~K at 105~GPa) \cite{Tian:2015-MgCH4}, H$_3$P$_{0.15}$S$_{0.85}$ (197~K at 200~GPa) \cite{Fan:2016-HPS-HSiP-HClS}, and CaYH$_{12}$ (258~K at 200~GPa) \cite{Liang:2019}. 
However, with notable exceptions  \cite{Shao:2019,Li:2020,Sun:2019,Zurek:2020b,Sun:2020a},
CSP searches were not carried out for a broad range of potential stoichiometries, and only a few compositions were considered.  

Herein, CSP searches using the \textsc{XtalOpt} evolutionary algorithm (EA) \cite{Zurek:2011a,avery2017xtalopt,avery2018xtalopt} 
coupled with Density Functional Theory (DFT) calculations were carried out to search for stable and intriguing metastable ternary hydrides (see the Supplementary Information, SI, for the computational details). Inspired by the high $T_c$ values predicted for CaH$_6$ \cite{wang2012superconductive}, as well as those computed \cite{Cui:2014a} and measured \cite{drozdov2015conventional} in H$_3$S, we focused on the Ca-S-H system.  EA runs were carried out on CaS and CaSH$_n$ ($n=$~1-4, 6, 8, 9, 12, 18, 20) with 1-4 formula units comprising the unit cell at 100, 150, 200, 250 and 300~GPa.  We found stable $n=2,6,20$ phases composed of a Ca-S lattice wherein the Ca ions were surrounded ``side-on'' by H$_2$ molecules that underwent H$_2$ $\sigma\rightarrow$ Ca $d$ donation, and Ca $d\rightarrow$ H$_2$ $\sigma^*$ back-donation. These phases were unlikely to be good superconductors. $P\bar{6}m2$ CaSH$_3$ contained CaH$_2$ layers, and two dimensional HS honeycomb sheets. Remarkably, just as in H$_3$S, the density of states at the Fermi level in $P\bar{6}m2$ CaSH$_3$ lies on a peak due to two van Hove singularities, thereby increasing its $T_c$, which was estimated to be as high as \edit{94-101~K at 128~GPa}.

\begin{figure*}
\begin{center}
\includegraphics[width=1.0\columnwidth]{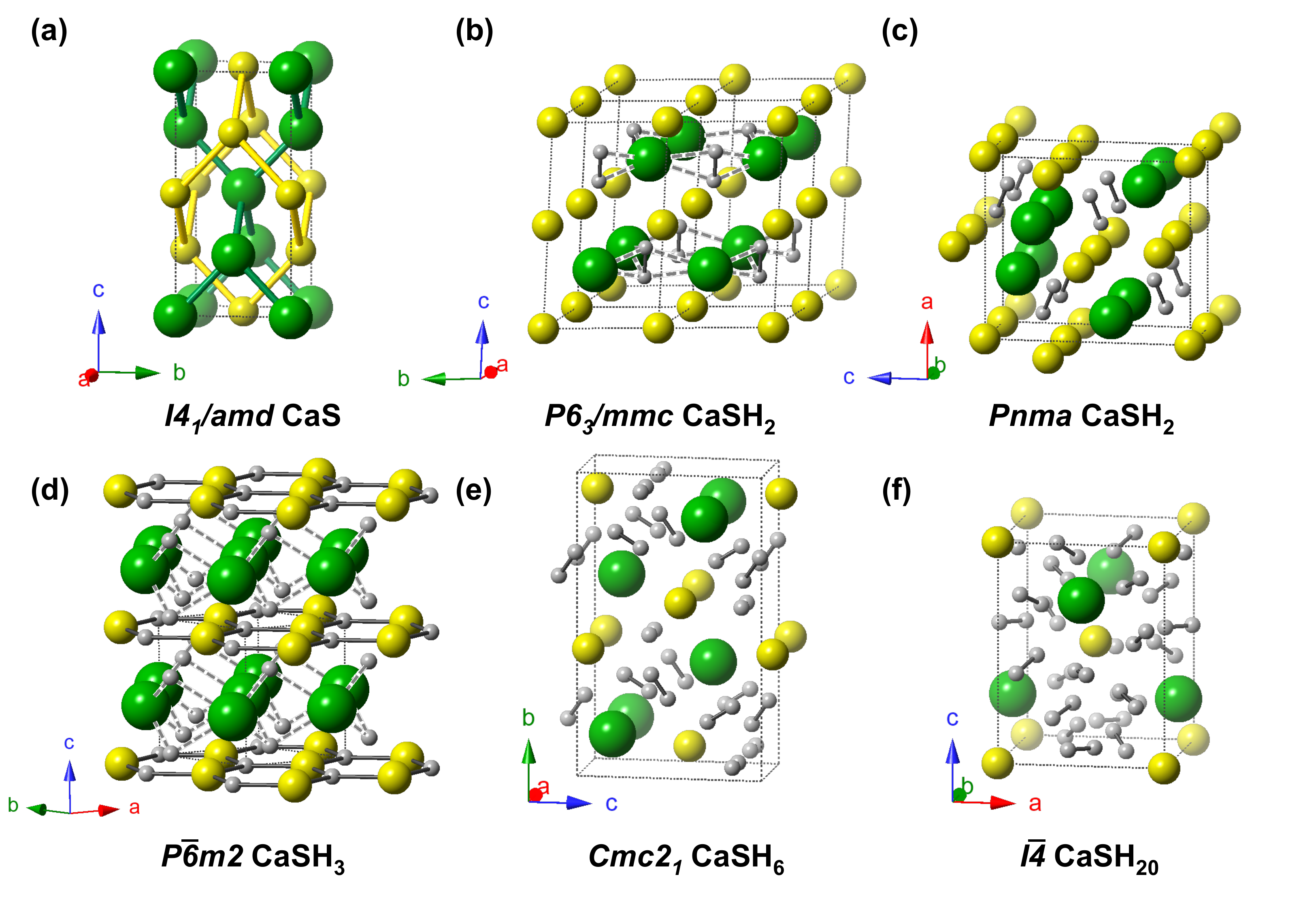}
\end{center}
\caption{Crystal structures of predicted CaSH$_n$ phases under pressure: (a) tetragonal $I4_1/amd$ CaS, (b) hexagonal $P6_3/mmc$ CaSH$_2$, (c) orthorhombic $Pnma$ CaSH$_2$, (d) hexagonal $P\bar{6}m2$ CaSH$_3$, (e) orthorhombic $Cmc2_1$ CaSH$_6$, and (f) tetragonal $I\bar{4}$ CaSH$_{20}$. Ca/S/H atoms are colored green/yellow/white. The solid lines connecting select hydrogen atoms indicate a bond, otherwise dashed lines are provided to guide the eye.
\label{fig:structures}}
\end{figure*}

CaS undergoes a $B1\rightarrow B2$ phase transition near 40~GPa \cite{luo1994structural}, and experiments have shown this structure is retained to 150~GPa \cite{ghandehari:2001}. Accordingly, our EA searches at 100 and 150~GPa identified the $B2$ phase as being the most stable. Above 170~GPa $I4_1/amd$ CaS, see Fig.\ \ref{fig:structures}(a), emerged as the global minimum, and phonon calculations showed it was  dynamically stable to at least 300~GPa (Figs.\ S5 and S7). $I4_1/amd$ CaS is composed of two interpenetrating Ca and S lattices that are both isotypic to a previously proposed phase of atomic metallic hydrogen \cite{Mcmahon:2011b}. Analysis of the electronic structure showed that the bonding is fully ionic, and PBE calculations suggested this phase may be a weak metal at 170~GPa with pressure induced band broadening increasing the number of bands that cross the Fermi level, $E_F$ (Fig.\ S6).

\begin{figure}[ht!]
\begin{center}
\includegraphics[width=1.0\columnwidth]{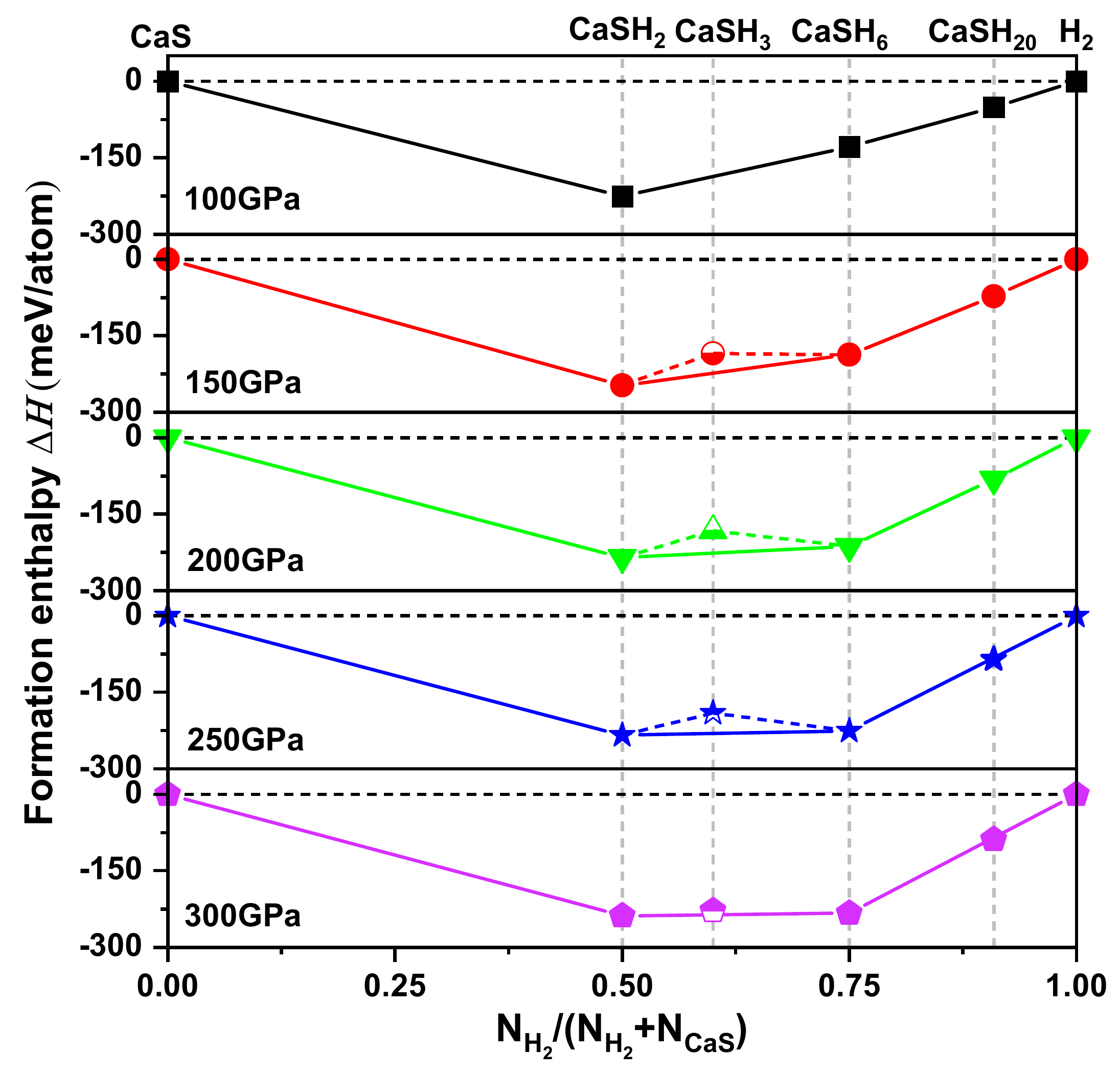}
\end{center}
\caption{$\Delta H$ for the reaction $(\frac{1}{2}\text{H}_2)_n+\text{CaS}\rightarrow \text{CaSH}_n$ versus the mole fraction of H$_2$ in the ternary as a function of pressure. For H$_2$ we used the enthalpy of the $C2/c$ (100-250~GPa) and $Cmca$-12 (300~GPa) phases \cite{pickard2007:H2}, and for CaS the  $Pm\bar{3}m$ (100-150~GPa) \cite{luo1994structural}, and $I4_1/amd$ (200-300~GPa) phases. Solid symbols lie on the convex hull (denoted by the solid lines), open symbols represent points above it. The ZPE-corrected convex hulls are given in Fig.\ S3.
\label{fig:2Dconvexhull}}
\end{figure}

Since this study is concerned with compounds having the CaSH$_n$ stoichiometry, which could be synthesized from CaS and H$_2$, we analyzed the 2D convex hulls that plot the enthalpy of formation, $\Delta H$, from these two starting materials between 100-300~GPa (Fig.\ \ref{fig:2Dconvexhull}). The stoichiometries that lay on the 2D hull within the whole pressure range considered were CaSH$_2$, CaSH$_6$ and CaSH$_{20}$. Generally speaking the same species comprised the 3D hull (Fig.\ S2), the only exception being that  the reaction $\text{CaSH}_{20}\rightarrow \text{CaH}_{12}+\text{H}_3\text{S}+\text{2.5H}_2$ was favored by 7~meV/atom at 250~GPa, and $\text{CaSH}_{20}\rightarrow \text{CaH}_{9}+\text{H}_3\text{S}+\text{4H}_2$ was favored by 23~meV/atom at 300~GPa, within the static lattice approximation. When the zero-point-energy (ZPE) was taken into consideration (Fig.\ S3), CaSH$_{20}$ rose by no more than 4~meV/atom above the 2D hull at 100-250~GPa. By \edit{128~GPa} the CaSH$_3$ stoichiometry emerged as being metastable. When ZPE corrections were considered it lay 3-28~meV/atom above the 150-250~GPa 2D hulls, and it fell on the 300~GPa hull. \edit{Recent research on binary hydrides under pressure have shown that enthalpies calculated within the static-lattice approximation cannot be employed as the only predictors of synthesizability. Experimental parameters including the choice of the precursors and temperatures employed can be tuned to obtain various kinetically stable products. For example, the preparation temperature was the key variable controlling if the H$_2$S sample employed by Drozdov and co-workers persisted, or decomposed into H$_3$S \cite{drozdov2015conventional}. Moreover,  DFT calculations performed with classical nuclei have shown that the enthalpy of the $Fm\bar{3}m$ LaH$_{10}$ phase is higher than that of other less symmetric structural alternatives, and it is dynamically unstable within the pressure range that superconductivity has been measured \cite{somayazulu2019evidence, drozdov2019superconductivity}. However, a recent theoretical study illustrated that anharmonic and quantum nuclear effects stabilize this phase, yielding computed $T_c$s  consistent with experiment \cite{Errea:2020a}.}
Therefore, both the stable and low-lying metastable Ca-S-H phases illustrated in Fig.\ \ref{fig:structures} will be considered herein. 

Between 100-160~GPa CaSH$_2$ adopts the $P6_3/mmc$ symmetry structure, composed of honeycomb layers wherein each Ca is coordinated to three H$_2$ molecules in a ``side-on'' fashion, and the S atoms are found in layers above and below the centers of the hexagons (Fig.\ \ref{fig:structures}(b)). $Pnma$ CaSH$_2$, stable at higher pressures, can be derived via a distortion of the hexagonal phase that reduces its volume (19.68~\AA{}$^3$/f.u.\ vs.\ 20.42~\AA{}$^3$/f.u.\ at 300~GPa) (Fig.\ \ref{fig:structures}(c)). Throughout the whole pressure range studied CaSH$_6$ and CaSH$_{20}$ adopted the $Cmc2_1$ and  $I\bar{4}$ spacegroups, respectively (Fig.\ \ref{fig:structures}(e) and (f)). Both were also characterized by Ca surrounded ``side-on'' by H$_2$ molecules. The ``side-on'' coordination of Ca by H$_2$ was first observed in an $I4/mmm$ symmetry CaH$_4$ phase that was theoretically predicted \cite{wang2012superconductive} prior to its synthesis \cite{Zurek:2018b}. This phase can be thought of as (CaH$_2$)(H$_2$)$_2$ whose H-H bond lengths were computed to be somewhat longer than those within solid H$_2$ at the same pressure. Detailed theoretical calculations showed that the increased bond length can be explained via a Kubas-like \cite{Kubas:1984} mechanism wherein H$_2$ donates $\sigma$ electrons to a vacant Ca d-orbital, and back-bonding from an occupied Ca d-orbital to the H$_2$ $\sigma^*$ further weakens the H-H bond \cite{Zurek:2018b}. \edit{Pressure is known to induce s $\rightarrow$ d transfer in Ca  \cite{maksimov:2005} so, like in CaH$_4$, the occupied Ca states in CaSH$_n$ are primarily d-like  (Fig.\ S12).} To further investigate if this mechanism occurs within the ternaries we calculated the negative of the crystal orbital Hamilton population integrated to the Fermi level (-ICOHP) for select atom pairs since it is a measure of their bond strength (Table \ref{tab:bond}).

Most of the H-H bond lengths in $P6_3/mmc$ and $Pnma$ CaSH$_2$, $Cmc2_1$ CaSH$_6$ and $I\bar{4}$ CaSH$_{20}$ are longer than those computed for molecular H$_2$ at the same pressure. Table \ref{tab:bond} shows that  the H-H bond strength is smaller in these phases than in H$_2$ molecules that adopt the CaSH$_n$ structure, but where the Ca and S atoms have been removed. The reason for this is that  H$_2$ $\sigma\rightarrow$ Ca $d$ donation, and Ca $d \rightarrow$ H$_2$ $\sigma^*$ back-donation \edit{(Fig.\ S13)} weakens the H-H bond in these ternaries, in-line with the findings for CaH$_4$ \cite{Zurek:2018b}.   
The Bader charges illustrate that S becomes negatively charged because of electron transfer from Ca, and accordingly the interaction between these two atoms, as gauged by the -ICOHPs, is non-negligible, and even larger than in CaS at comparable pressures. The H-S bond strength is of a similar magnitude as the Ca-S interaction, and the markedly smaller H-Ca interaction is on par with results obtained previously for CaH$_4$ \cite{Zurek:2018b}.

\begin{table*}
    \centering
    \def\arraystretch{1.0}
    \caption{Distances between select atom pairs in $Pm\bar{3}m$ and $I4_1/amd$ CaS, $R3m$ and $Im\bar{3}m$ H$_3$S, $I4/mmm$ CaH$_4$, $P6_3/mmc$ and $Pnma$ CaSH$_2$, $Cmc2_1$ CaSH$_6$, $I\bar{4}$ CaSH$_{20}$ and their corresponding crystal orbital Hamilton populations integrated to the Fermi level (-ICOHPs) at various pressures. The -ICOHP for a hypothetical lattice of H$_2$ molecules where the Ca and S atoms are deleted are provided in parentheses. The H-H distance in solid molecular H$_2$ at 150, 200 and 300~GPa is 0.742, 0.745 and 0.756~\AA{}, with corresponding -ICOHPs of 6.74, 6.62, and 6.21~eV/bond.}
     \setlength{\tabcolsep}{3mm}{        
       \begin{tabular}{c c c c c l}
\hline
\hline
System & Space group & Pressure (GPa)& Atom Pairs & Distance (\AA{})  & -ICOHP (eV/bond) \\
\hline
CaS      & $Pm\bar{3}m$ & 150 & Ca-S & 2.419 & 0.40 \\
         & $I4_1/amd$  & 200 & Ca-S & 2.336 & 0.10 \\
         &              & 250 & Ca-S & 2.295 & 0.13 \\
         &              & 300 & Ca-S & 2.262 & 0.28 \\
\hline         
H$_3$S   & $R3m$ & 150 & H-S & 1.458 & 4.16 \\
         & $Im\bar{3}m$ & 200 & H-S & 1.492  & 3.51 \\
         &            & 300 & H-S & 1.433 & 3.73 \\
\hline

CaH$_4$ & $I4/mmm$ & 120 & H-H & 0.811 & 4.15 \\
                &           &     & H$_\text{atomic}$-Ca & 1.901 & 0.24 \\
                &           &     & H$_\text{molecular}$-Ca & 2.048 & 0.11 \\
                &           &     & H$_\text{molecular}$-Ca & 2.124 & 0.16 \\
\hline
CaSH$_2$ & $P6_3/mmc$ & 150 & H-H &0.811 &  3.92 (5.31)\\

         &            &     & H-S & 2.117 &  0.38 \\

         &            &     & Ca-S & 2.340 & 0.65 \\
         
                  &            &     & H-Ca & 1.898 & 0.09 \\
\hline
CaSH$_2$ & $Pnma$ & 300   & H-H  & 0.814 &  3.61 (5.51) \\

         &            &     & H-S  & 1.745 &  0.84 \\
         
         &          &     & Ca-S & 2.202  & 0.77 \\
         
                  &          &     & H-Ca & 1.766  & 0.09 \\
\hline
CaSH$_3$ & $P\bar{6}m2$ & 300 & H-S  & 1.531 & 2.22 \\
          &         &       & H-S   & 1.704  & 1.01 \\  
           &          &     & Ca-S & 2.289  & 0.44 \\
           &         &        &  H-Ca & 1.803 & 0.18 \\
\hline
CaSH$_6$ & $Cmc2_1$ & 300 & H-H & 0.770 &  4.43 (5.41) \\

         &          &     & H-H & 0.776 & 4.25 (5.16) \\

         &          &     & H-S & 1.694 &  1.25 \\
         
         &          &     & Ca-S & 2.276 & 0.72 \\   
          
         &          &     & Ca-S & 2.288 & 0.73 \\
         
        &          &     & H-Ca & 1.743 &  0.12 \\
\hline
CaSH$_{20}$ & $I\bar{4}$ & 200 & H-H & 0.736 &  5.50 (6.29)\\
          &         &     & H-H  & 0.750 & 5.26 (6.01)\\
          &         &     & H-H & 0.760 & 4.89 (6.02) \\
          &         &     & H-S & 1.744  & 1.33 \\
        &          &     & Ca-S & 2.592  & 0.50 \\
                  &         &     & H-Ca & 1.863 &  0.15 \\
\hline                
\hline
\end{tabular}
\label{tab:bond}}
\end{table*}

A number of hydrogenic motifs have been found in binary hydrides of the alkali, alkaline and rare earth metals under pressure \cite{Zurek:2018m}. The presence of H$_2^{\delta-}$ units, whose bonds were slightly elongated relative to the elemental phase, typically yielded good metals with moderate $T_c$s \cite{Zurek:2018m}. Phases where other species, such as H$^-$ or H$_3^-$, were present along with H$_2$, were usually insulators or weak metals \cite{Zurek:2016j,Zurek:2016d}. Similar to the results obtained for (CaH$_2$)(H$_2$)$_2$ \cite{Zurek:2018b}, we speculated that the ternary hydrides, whose formulae can be written as (Ca$^{2+}$S$^{2-}$)(H$_2$)$_n$ in the full ionic picture, were unlikely to have an electronic structure that was conducive towards superconductivity. Band structure calculations carried out with the PBE functional suggested that $Pnma$ CaSH$_2$ at 300~GPa, and $I\bar{4}$ CaSH$_{20}$ at 200~GPa might be weak metals (Fig.\ S9). However, HSE06 screened hybrid functional calculations, which provide better estimates of the band gap, yielded gaps of $\sim$0.4~eV (Fig.\ S10). This suggests that ternary hydrides containing an electropositive Group I, II or III element, combined with a $p$-block element are unlikely to be good superconductors if the only hydrogenic species found within them are H$_2$ units.

Three CaSH$_3$ lattices emerged as being the most stable in our EA searches at different pressures (Fig.\ S8). $P\bar{6}m2$ CaSH$_3$ (Fig.\ \ref{fig:structures}(d)) lay on the 300~GPa 2D hull including the ZPE (Fig.\ S3).  At pressures smaller than $\sim$240~GPa other phases had lower enthalpies, but $P\bar{6}m2$ CaSH$_3$ remained dynamically stable to \edit{128}~GPa. \edit{Molecular dynamics simulations (Section S18) at 240 and 140~GPa showed that this phase is thermally stable, as it did not undergo any structural reconstructions, thereby suggesting that it can be quenched to lower pressures}. $P\bar{6}m2$ CaSH$_3$ did not contain any H$_2$ molecules, and it was composed of honeycomb SH layers wherein each S atom was bonded to three H atoms (and vice versa), and CaH$_2$ layers, so its formula can be written as (CaH$_2$)(SH). $Im\bar{3}m$ H$_3$S can be described as two interpenetrating SH$_3$ perovskite sublattices, wherein each S is octahedrally coordinated by H, and each H is coordinated to two S atoms \cite{Cui:2014a}. The nearest neighbor H-S bond lengths in $P\bar{6}m2$ CaSH$_3$, 1.531~\AA{} at 300~GPa, are somewhat longer and weaker than those in SH$_3$ at the same pressure, 1.433~\AA{} (-ICOHPs of 2.22 vs.\ 3.73~eV/bond). Below we investigate if other characteristics of $P\bar{6}m2$ CaSH$_3$ bear a resemblance to those computed for superconducting H$_3$S.

\begin{figure}
\begin{center}
\includegraphics[width=0.7\columnwidth]{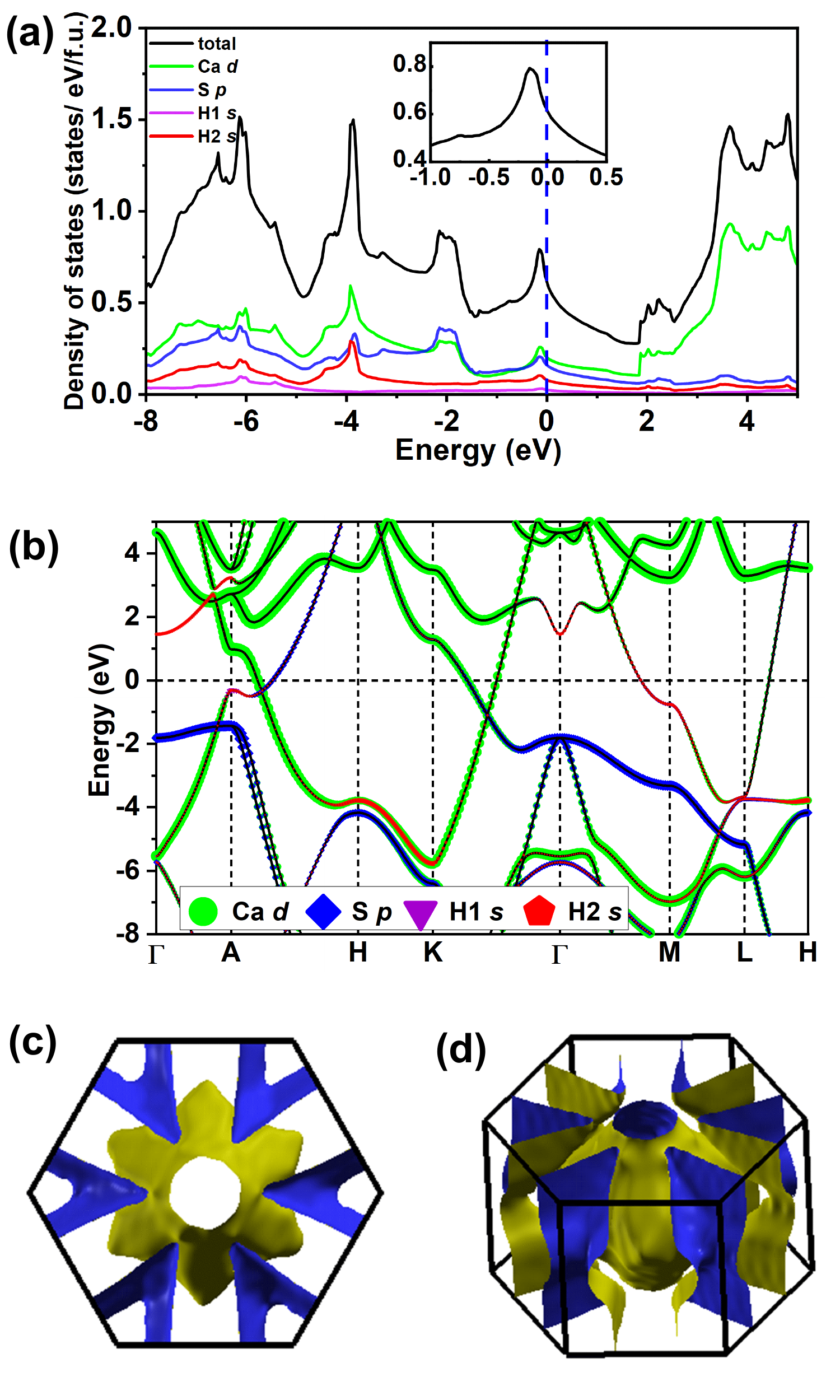}
\end{center}
\caption{The electronic structure of $P\bar{6}m2$ CaSH$_3$ at 250~GPa: (a) projected density of states (DOS), with the inset showing a blow up around the Fermi level, (b) fat bands illustrating the character arising from the \edit{H1/H2 $s$ states (also see Fig.\ S11)}, S $p$ states, and Ca $d$ states, and (c,d) Fermi surface -- top and side view. \edit{H1 atoms are in the ``SH'' layers, and H2 atoms are in the ``CaH$_2$'' layers  (see Fig.\ \ref{fig:structures}(d))}.
\label{fig:cash3banddos}}
\end{figure}

The projected density of states (DOS) at 250~GPa, and the corresponding band structure projected onto atomic orbitals (fat bands) of $P\bar{6}m2$ CaSH$_3$ are plotted in Fig.\ \ref{fig:cash3banddos}. They show that Ca $d$, S $p$ and H $s$ states are found to give the largest contributions to the DOS at $E_\text{F}$. \edit{There are twice as many hydrogen atoms in the CaH$_2$ layers as compared with the SH layers, so it is no surprise that the DOS is larger for the former than the latter. However, along the high symmetry lines crossing  $E_\text{F}$ the character is primarily due to the CaH$_2$ hydrogens, and the band just below $E_\text{F}$ at the $A$-point contains character from both (Fig.\ S11)}. Remarkably, the DOS at $E_\text{F}$ lies close to the top of a particularly sharp peak that is due to two van Hove singularities (vHs) separated by $\sim$64~meV (\edit{Fig.\ S24}). Because vHs increase the number of states at $E_\text{F}$ that can participate in the electron-phonon-coupling (EPC) mechanism, they are known to enhance the total coupling strength, $\lambda$, and in turn the $T_c$ in conventional superconductors. It is therefore not surprising that DFT calculations showed that $Im\bar{3}m$ H$_3$S and $Fm\bar{3}m$ LaH$_{10}$ exhibit such double-shaped vHs at $E_\text{F}$ separated by $\sim$300~meV  \cite{quan2016van} and $\sim$90~meV  \cite{liu2019microscopic}, respectively. The role that vHs play in increasing the $T_c$ in H$_3$S has been studied in great detail \cite{quan2016van,sano2016effect,Ortenzi:2016,Akashi:2020}. Between \edit{128}-300~GPa the vHs in CaSH$_3$ are pinned near $E_\text{F}$ (Fig.\ S14), and the Fermi surface remains nearly invariant \edit{(Figs.\ S20-S23)}. \edit{By 140~GPa}, however, a band that displays H s character rises above $E_\text{F}$ at the $A$-point, resulting in a Lifshitz transition. As shown in Table \ref{tab:Tc}, the Lifshitz transition is associated with a pronounced increase of the DOS at $E_\text{F}$, suggesting that $T_c$ may be largest near the onset of dynamic instability, which occurs near \edit{128~GPa (Fig.\ S25)}.

\begin{figure}
\begin{center}
\includegraphics[width=1.0\columnwidth]{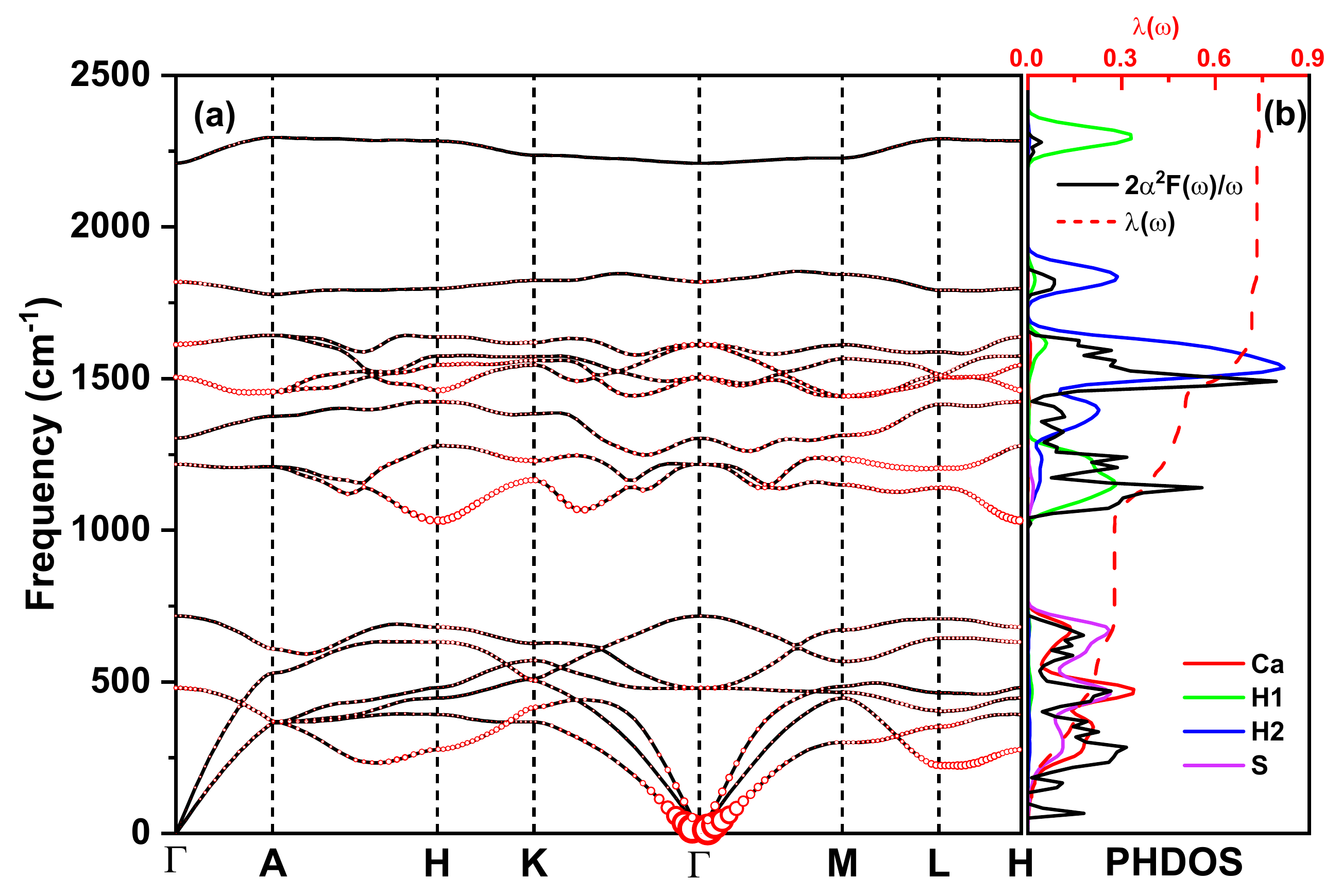}
\end{center}
\caption{Superconducting properties of $P\bar{6}m2$ CaSH$_3$ at 250 GPa: (a) phonon band structure where the radius of the circles is proportional to the electron-phonon coupling constant, $\lambda_{\textbf{q}\nu}$, for mode $\nu$ at wavevector $\textbf{q}$ (phonon linewidth), and (b) the Eliashberg spectral function (black line), electron phonon integral, $\lambda(\omega)$ (red dashed line), and projected phonon densities of states (PHDOS). H1 atoms are in the ``SH'' layers, and H2 atoms are in the ``CaH$_2$'' layers  (see Fig.\ \ref{fig:structures}(d)).
\label{fig:250G_elph}}
\end{figure}
    
Subsequently, we calculated the phonon band structures, projected phonon DOS, Eliashberg spectral function, $\alpha^2F(\omega)$, and the EPC integral, $\lambda(\omega)$, for $P\bar{6}m2$ CaSH$_3$  within the range of its dynamic stability (Figs.\ \ref{fig:250G_elph}, S25 and S26).  At 250~GPa two gaps separate the phonon spectrum into three regions: the lower frequency modes are mainly associated with the motions of the heavier calcium and sulfur atoms, while the hydrogens in the SH layers (H1) and those in the CaH$_2$ layers (H2) contribute to the intermediate region, and the higher frequency modes are associated with the H1 atoms, with these regions contributing 37.6, 59.4, and 3\% to the total $\lambda$, respectively. Most of the phonon modes harden up to 300~GPa resulting in an increased logarithmic average of phonon frequencies, $\omega_\text{log}$.  On the other hand, the EPC increases with decreasing pressure, attaining a maximum value of $\lambda=$~1.64 at 128~GPa. $T_c$ was estimated using the Allen-Dynes modified McMillan equation \cite{allen1975McMllanequation}, along with typical values of the renormalized Coulomb repulsion, $\mu^*=0.1-0.13$, and it was found to fall in the range of 36-46~K at 250~GPa, increasing to 66-75~K by 140~GPa (Table \ref{tab:Tc}). \edit{It is well-known that for strongly coupled systems, $\lambda\gtrsim1.5$, this method underestimates $T_c$. Numerically solving the Eliashberg equations provides a more accurate estimate, resulting in values as high as 94-101~K at 128~GPa.}

The phonon linewidths illustrate that many modes in the low and intermediate frequency regions contribute towards $\lambda$.  A set of modes around, but not at, $\Gamma$ along the $K-\Gamma -M$ path stand out as having particularly large linewidths. Their frequencies are degenerate with one of the translational modes that corresponds to displacement of the 2D layers in the $xy$ plane.  Visualization of two arbitrarily chosen modes on the $K-\Gamma$ and $\Gamma-M$ paths  with  large linewidths  show that all of the atoms move in the $xy$ plane. Both modes are reminiscent of planewaves, but locally the motion along $K-\Gamma$ resembles rocking, and the one along $\Gamma-M$ scissoring. The phonon band structure at \edit{128~GPa} is not too different from the one at 250~GPa, except that the division into three separate regions is not so obvious because of the softening of the modes in the intermediate regime (Figs.\ S25 and S26). In addition to the modes along the $K-\Gamma -M$ high symmetry lines softened modes along the $A-H$ and $L-H$ paths also possess a large linewidth. \edit{At lower pressures the latter become imaginary, and the phase becomes dynamically unstable.}

Since $P\bar{6}m2$ CaSH$_3$ can be thought of as 2D SH layers sandwiched between CaH$_2$ sheets, it is instructive to compare its predicted superconducting properties with similar systems. The maximum EPC parameter calculated for CaSH$_3$, \edit{$\lambda=1.64$}, is lower than that of the strongly coupled H$_3$S phases \cite{Cui:2014a}, as is $\omega_\text{log}$ resulting in a $T_c$ that is about a factor of \edit{two} smaller. The EPC in a family of superconducting ternary hydrides based on an H$_3$S lattice intercalated by CH$_4$ ranged from $\lambda=1.06-3.64$, and $T_c$, as calculated using the Allen-Dynes modified McMillan equation with $\mu^*=0.1$, fell between 98-156~K \cite{Zurek:2020b,Sun:2020a}. Whereas the DOS at $E_\text{F}$ in CH$_4$-SH$_3$ only contained contributions from the H$_3$S framework, in CaSH$_3$ substantial Ca $d$-character was also present. Therefore, despite the presence of vHs near $E_F$ in both CaSH$_3$ and H$_3$S, the microscopic mechanism of superconductivity in these two phases is subtly different, whereas only the H$_3$S lattice contributes to the superconductivity in CH$_4$-SH$_3$.

\begin{table}
    \centering
    \def\arraystretch{1.0}
    \caption{Superconducting parameters for $P\bar{6}m2$ CaSH$_3$ at various pressures: electron-phonon coupling parameter ($\lambda$), logarithmic average of phonon frequencies ($\omega_{\text{log}}$) and estimated superconducting critical temperature ($T_c$) for values of the Coulomb pseudopotential, $\mu^*$, of 0.1 and 0.13 using the Allen-Dynes modified McMillan equation \cite{allen1975McMllanequation} \edit{(and numerically solving the Eliashberg equations \cite{eliashberg:1960})}. The density of states at the Fermi level, $N(E_\text{F})$, is also provided in units of states/eV/f.u.}
        \begin{tabular}{c c c c c c}
\hline
\hline
Pressure (GPa) & $\lambda$ & $\omega_{\text{log}}$ (K)  & T$_c^{\mu^*=~0.1}$ (K) & T$_c^{\mu^*=~0.13}$ (K) & $N(E_f)$ \\
\hline
128 &  1.64 & 629.6 & 78 \edit{(101)} & 71 \edit{(94)} & 0.74 \\

140 & 1.18 & 861.2 & 75 \edit{(91)} &  66 \edit{(82)} & 0.72  \\

200 & 0.82 & 1102.1 & 54 &  44 & 0.64 \\

250 & 0.74 & 1170.5 & 46 &  36 & 0.61 \\

300 & 0.70 & 1211.0 & 42 &  32 & 0.59 \\

\hline
\end{tabular}
\label{tab:Tc}
\end{table}

In conclusion, evolutionary searches have identified five hitherto unknown phases with unique stoichiometries in the ternary Ca-S-H system that may be synthesizable at pressures of 100-300~GPa. $P6_3/mmc$ CaSH$_2$, $Pnma$ CaSH$_2$, $Cmc2_1$ CaSH$_6$, and $I\bar{4}$ CaSH$_{20}$, which comprise the 2D convex hull, contain Ca, S and H$_2$ molecules. The H-H distances in these semiconductors are elongated as compared to molecular H$_2$ because of H$_2$ $\sigma\rightarrow$ Ca $d$ donation, and Ca $d$~$\rightarrow$ H$_2$ $\sigma^*$ back-donation via a Kubas-like mechanism.  Moreover, we also predict a metallic $P\bar{6}m2$ CaSH$_3$ phase composed of 2D honeycomb HS layers separated by layers of CaH$_2$ that is metastable between \edit{128}-300~GPa. The DOS at $E_\text{F}$, which exhibits Ca $d$, S $p$ and H $s$ character, lies near a peak that is due to two van Hove singularities (vHs). Remarkably, $E_\text{F}$ remains pinned at the vHs, and the Fermi surface is nearly invariant within the range of dynamic stability. \edit{By 140}~GPa a new band rises above $E_\text{F}$ resulting in a Lifshitz transition that markedly increases the DOS at $E_\text{F}$, and concomitantly the EPC. Estimates of the $T_c$ via the Allen-Dynes modified McMillan equation range from 32-42~K at 300~GPa, \edit{whereas numerically solving the Eliashberg equations suggests $T_c$ may be as high as 94-101~K at 128~GPa.} Our findings provide a microscopic understanding of the superconducting mechanism in $P\bar{6}m2$ CaSH$_3$, and inspire the search for superconductivity in phases composed of \edit{honeycomb HX (X=S, Se, Te), and MH$_2$ (M=Mg, Ca, Sr, Ba) layers. For example, exploratory calculations (Section S19) shown that an isotypic CaSeH$_3$ phase is dynamically stable at pressures as low as 110~GPa. Its band structure resembles that of $P\bar{6}m2$ CaSH$_3$, except bands that are unoccupied in CaSH$_3$ cross $E_F$ along the $\Gamma-A$ and $K-\Gamma-M$ high symmetry lines. We are looking forward to future work studying the effect of the chemical substitution on the stability and superconducting properties of phases isotypic with $P\bar{6}m2$ CaSH$_3$ under pressure.} \\

\textbf{Acknowledgements:}
This material is based upon work supported by the U.S.\ Department of Energy, Office of Science, Fusion Energy Sciences under Award No.\ DE-SC0020340 (Y.Y.). T.B., N.G., and X.W. acknowledge the U.S. National Science Foundation (DMR-1827815) for financial support. Calculations were performed at the Center for Computational Research at SUNY Buffalo \cite{ccr}. We thank Russell Hemley for helpful discussions. \\

\textbf{Supporting Information:}
The Supporting Information is available free of charge on the ACS Publication website. It includes the computational details, illustrations of the 3D and 2D convex hulls, structural parameters, relative enthalpies, electronic band structures and densities of states, phonon band structures, electron localization functions, and Bader charges of select CaS and CaSH$_n$ ($n=2,3,6,20$) phases, as well as the Fermi surface, \edit{isoenergy contours, snapshots of MD simulations, and Eliashberg spectral function of $P\bar{6}m2$ CaSH$_3$ at different pressures}.


\providecommand{\latin}[1]{#1}
\makeatletter
\providecommand{\doi}
  {\begingroup\let\do\@makeother\dospecials
  \catcode`\{=1 \catcode`\}=2 \doi@aux}
\providecommand{\doi@aux}[1]{\endgroup\texttt{#1}}
\makeatother
\providecommand*\mcitethebibliography{\thebibliography}
\csname @ifundefined\endcsname{endmcitethebibliography}
  {\let\endmcitethebibliography\endthebibliography}{}

\end{document}